\documentclass[11pt,a4paper]{article}

\usepackage[utf8]{inputenc}
\usepackage{jcappub}
\usepackage{booktabs}
\usepackage{microtype}
\hypersetup{hidelinks}

\title{Rainbow McVittie Horizons in an Expanding Universe}
\author[a]{Amani Ashour,}
\author[b,c]{and Ahmed Farag Ali}
\affiliation[a]{Faculty of Informatics and Communication Engineering, Yarmouk Private University,
Jabab, Damascus--Daraa Highway, Syria}
\affiliation[b]{Essex County College,
303 University Avenue, Newark, New Jersey 07102, USA}
\affiliation[c]{Department of Physics, Faculty of Science, Benha University,
 Benha 13511, Qalubiya, Egypt}

\abstract{
We derive the Friedmann dynamics and apparent-horizon thermodynamics of a spatially flat McVittie spacetime in Gravity's Rainbow and construct a consistent non-accreting perfect-fluid sector with an effective rainbow Hubble rate. The Einstein equations, trapping horizon, surface gravity, Misner--Sharp energy, unified first law, and Clausius relation form a closed thermodynamic system reproducing the Friedmann dynamics. We then introduce a low-energy rainbow deformation and constrain it using 32 cosmic-chronometer measurements and 1580 Pantheon+ light curves with the full covariance matrix. Combined with the Planck matter-density prior, the fit gives $H_0=68.75\pm2.63~\mathrm{km\,s^{-1}\,Mpc^{-1}}$, $\Omega_m=0.3157\pm0.0070$, and $\epsilon=0.0106^{+0.0234}_{-0.0231}$. The resulting constraints establish the observational viability of small late-time rainbow corrections and place the general-relativistic limit within the preferred parameter region.

}

\begin{document}
\maketitle

\section{Introduction}

The study of cosmology within the framework of General Relativity often relies on the Friedmann equations \cite{friedmann1922,friedmann1924}, which describe the dynamics of a homogeneous and isotropic universe. Traditionally, these equations are derived from the Einstein field equations assuming the Friedmann--Lema\^itre--Robertson--Walker (FLRW) metric \cite{lemaitre1927}.

A central mass embedded in an expanding universe is described naturally by the McVittie generalization of the FLRW spacetime \cite{mcVittie1933,nolan1998,kaloper2010,faraoni2012}. This metric allows one to investigate the interplay between local gravitational effects and the overall cosmic expansion.

Furthermore, thermodynamic formulations of gravitational dynamics connect spacetime evolution to entropy and energy fluxes at horizons \cite{jacobson1995,hayward1998}. In spherical symmetry, this framework gives a particularly direct relation between the Misner--Sharp energy, work density, energy-supply vector, and trapping-horizon dynamics.

Gravity's Rainbow was introduced as an energy-dependent extension of doubly special relativity in which the effective orthonormal frame, and therefore the metric probed by a particle, depends on the ratio $E/E_{\rm P}$ through two functions $f(E/E_{\rm P})$ and $g(E/E_{\rm P})$ \cite{magueijo2004}. Early developments established two principal directions that remain central to the present work. Rainbow black-hole thermodynamics was developed through energy-dependent temperature and entropy relations \cite{galan2006,lingli2007}, while rainbow cosmology generalized the Friedmann description to energy-dependent geometries \cite{ling2007}.

The cosmological program includes nonsingular FRW solutions generated by rainbow functions associated with modified dispersion relations \cite{awad2013}, together with formulations connecting rainbow frames to scale-invariant primordial fluctuations and dimensional reduction \cite{amelino2013}. Further developments relate Gravity's Rainbow to Ho\v{r}ava--Lifshitz gravity \cite{garattini2015}, Gauss--Bonnet rainbow cosmology \cite{hendi2016gb}, and the effective Friedmann dynamics of loop quantum cosmology \cite{gorji2017}. These studies establish an extensive cosmological setting in which energy-dependent geometry modifies early-universe dynamics and motivates controlled low-energy limits.

The thermodynamic branch developed in parallel. Rainbow-modified Schwarzschild thermodynamics yields finite black-hole remnants for representative modified dispersion relations \cite{ali2014}, with extensions to black rings \cite{aliblackrings2014}, general rotating and charged black objects \cite{alifaizal2015remnant}, and collider-scale implications \cite{alifaizal2015lhc}. The same framework has been extended to black $p$-branes, including calculations of temperature, entropy, heat capacity, and remnant structure \cite{ashour2016}. Rainbow deformations have also been applied to neutron-star equilibrium and stability \cite{hendi2016tov} and to black-hole phase structure \cite{feng2017}. Together, this literature motivates an energy-dependent McVittie construction that unifies cosmological expansion, local gravitating structure, and horizon thermodynamics within one spherical geometry.

We introduce a prescribed energy history $E(t)$ and use the equivalent notation $f(t)$ and $g(t)$. The variables $d\tau=dt/f$ and $b=a/g$ isolate the kinematical reparametrization carried by one such history and preserve the standard non-accreting McVittie matter content. This exact reduction also fixes the logical scope of the construction: a single prescribed history contains the geometry of standard McVittie, whereas invariant rainbow information can arise from comparisons among energy sectors, an independently specified evolution law for $E$, an energy-dependent coupling, or modified matter propagation. The observational response introduced below is consequently an explicit phenomenological closure whose assumptions remain separate from the exact geometric derivation.

\section{Modified Friedmann Equations for the McVittie Metric in GR}

The McVittie metric for a flat spatial section ($k=0$) is given by \cite{mcVittie1933}
\begin{equation}\label{mcmetric}
 ds^2=-\left(\frac{1-\mu}{1+\mu}\right)^2dt^2+a(t)^2(1+\mu)^4\left(dr^2+r^2d\Omega^2\right),
\end{equation}
where
\begin{equation}
 \mu=\frac{m}{2a(t)r}, \qquad H=\frac{\dot a}{a},
\end{equation}
and $m$ is the constant McVittie mass parameter. We use $G=c=\hbar=k_B=1$ throughout.

The isotropic chart covers the exterior region $0\leq\mu<1$, equivalently $R>2m$. We take an expanding sector, $H>0$, when discussing the two positive apparent-horizon branches.

The areal radius is
\begin{equation}\label{Rclass}
 R=a(t)r(1+\mu)^2,
\end{equation}
and it is useful to define
\begin{equation}\label{chiDef}
 \chi\equiv\frac{1-\mu}{1+\mu}=\sqrt{1-\frac{2m}{R}}.
\end{equation}
The two-dimensional metric normal to the symmetry spheres is
\begin{equation}
 h_{ij}=\mathrm{diag}\left(-\chi^2,\,a^2(1+\mu)^4\right).
\end{equation}
For a comoving perfect fluid,
\begin{equation}
 T_{\mu\nu}=(\rho+p)u_\mu u_\nu+pg_{\mu\nu},\qquad
 u^\mu=\left(\chi^{-1},0,0,0\right).
\end{equation}

For the non-accreting McVittie solution, the density is homogeneous, $\rho=\rho(t)$, while the pressure is generally inhomogeneous, $p=p(t,r)$. The independent Einstein equations are
\begin{align}
 H^2&=\frac{8\pi}{3}\rho,\label{frclassic1}\\
 \dot H&=-4\pi(\rho+p)\chi
 =-4\pi(\rho+p)\sqrt{1-\frac{2m}{R}}.\label{frclassic2}
\end{align}
Equivalently,
\begin{equation}
 8\pi p=-3H^2-\frac{2\dot H}{\chi}.
\end{equation}
The covariant conservation law $\nabla_\mu T^{\mu\nu}=0$ gives
\begin{equation}\label{continuityClass}
 \dot\rho+3H\chi(\rho+p)=0,
\end{equation}
which is exactly consistent with differentiating Eq.~\eqref{frclassic1} and using Eq.~\eqref{frclassic2}. In the limit $m\to0$, one has $\chi\to1$ and the usual flat-FLRW equations are recovered.

\subsection{Apparent horizon of the McVittie metric}

For a spherically symmetric spacetime, the apparent horizon is determined by \cite{hayward1996}
\begin{equation}
 h^{ab}\partial_aR\partial_bR=0.
\end{equation}
From Eq.~\eqref{Rclass},
\begin{equation}
 \partial_tR=HR\chi,\qquad \partial_rR=a(1-\mu^2),
\end{equation}
and therefore
\begin{equation}\label{horizonClass}
 1-\frac{2m}{R_A}-H^2R_A^2=0,
 \qquad\Longleftrightarrow\qquad
 R_A-2m-H^2R_A^3=0.
\end{equation}
Two positive roots exist when $0<3\sqrt{3}\,mH<1$: the black-hole branch obeys $2m<R_{Ab}<3m$, and the cosmological branch obeys $R_{Ac}>3m$. They merge at $R_A=3m$ when $3\sqrt{3}\,mH=1$. The analysis below applies to either nondegenerate branch and states explicitly when division by $3m-R_A$ is used.

For dynamical spherical spacetimes, the dynamical-horizon surface gravity is
\begin{equation}
 \kappa=\frac{1}{2\sqrt{-h}}\partial_i\left(\sqrt{-h}\,h^{ij}\partial_jR\right).
\end{equation}
Evaluating it at a nondegenerate McVittie apparent horizon gives
\begin{equation}\label{kappaClass}
 \kappa_A=\frac{1}{R_A}\left(\frac{3m}{R_A}-1\right)
 -\frac{\dot H\,R_A}{2\sqrt{1-2m/R_A}}.
\end{equation}
Using the time derivative of Eq.~\eqref{horizonClass}, this is equivalent to
\begin{equation}
 \kappa_A=\frac{1}{R_A}\left(\frac{3m}{R_A}-1\right)
 \left(1-\frac{\dot R_A}{2HR_A\sqrt{1-2m/R_A}}\right).
\end{equation}
The physical Hawking--Kodama temperature is taken to be
\begin{equation}\label{Tclass}
 T_A=\frac{|\kappa_A|}{2\pi},
\end{equation}
where the absolute value defines the positive Hawking--Kodama temperature associated with either trapping-horizon orientation.

The Misner--Sharp energy is
\begin{equation}
 E=\frac{R}{2}\left(1-h^{ab}\partial_aR\partial_bR\right)
 =m+\frac{1}{2}H^2R^3
 =m+\frac{4\pi}{3}\rho R^3.
\end{equation}
Hence, on any apparent-horizon branch satisfying Eq.~\eqref{horizonClass},
\begin{equation}
 E_A=\frac{R_A}{2},\qquad S_A=\frac{A_A}{4}=\pi R_A^2,
\end{equation}
with $A_A=4\pi R_A^2$. The quantity $E_A=R_A/2$ is the Misner--Sharp energy evaluated on the trapping horizon, while $m$ remains the constant McVittie central-mass parameter.

\section{Consistent Time-Dependent Rainbow Parametrization}

The rainbow deformation used here follows the standard energy-dependent frame prescription developed in black-hole and cosmological applications \cite{magueijo2004,ling2007,awad2013,ali2014,ashour2016}. The direct ansatz
\begin{equation}\label{naiveMetric}
 ds^2=-\frac{\chi^2}{f(t)^2}dt^2+\frac{a(t)^2}{g(t)^2}(1+\mu)^4\left(dr^2+r^2d\Omega^2\right),
 \qquad \mu=\frac{m}{2ar},
\end{equation}
defines an areal radius $\widetilde R=R/g$ and therefore the effective mass relation
\begin{equation}
 \chi^2=1-\frac{2m}{g\widetilde R}=1-\frac{2M_{\rm eff}(t)}{\widetilde R},
 \qquad M_{\rm eff}=\frac{m}{g(t)}.
\end{equation}
The areal-radius geometry therefore carries the time-dependent effective mass $M_{\rm eff}=m/g(t)$. Since $\dot\mu=-H\mu$, the logarithmic derivative of the isotropic spatial factor satisfies
\begin{equation*}
 \frac{d}{dt}\ln\!\left[\frac{a}{g}(1+\mu)^2\right]
 =H\chi-G_g,
 \qquad G_g\equiv\frac{\dot g}{g}.
\end{equation*}
For zero shift, the isotropic extrinsic curvature is therefore
\begin{equation}\label{directExtrinsic}
 K^i{}_j=-f\left(H-\frac{G_g}{\chi}\right)\delta^i{}_j
 =-f\left[H-G_g\left(\frac{1+\mu}{1-\mu}\right)\right]\delta^i{}_j,
\end{equation}
where we use $K_{ij}=-(2N)^{-1}\partial_t\gamma_{ij}$ with lapse $N=\chi/f$. Thus the factor found by direct calculation is precisely $1/\chi=(1+\mu)/(1-\mu)$. Its radial derivative gives the explicit momentum-constraint source
\begin{equation}\label{naiveMomentum}
 D_j\!\left(K^j{}_r-\delta^j{}_rK\right)
 =-2\partial_rK^r{}_r
 =\frac{4fG_g\mu}{r(1-\mu)^2}.
\end{equation}
It vanishes for $m=0$ or $\dot g=0$ and remains nonzero for $m\dot g\neq0$. The linear divergence of $G_g/\chi$ in Eq.~\eqref{directExtrinsic} produces the quadratic divergence of its radial gradient in Eq.~\eqref{naiveMomentum}. Equation~\eqref{naiveMetric} therefore requires a source with radial momentum density in that regime and lies outside the non-accreting perfect-fluid McVittie class.

The non-accreting perfect-fluid McVittie sector is obtained by defining
\begin{equation}\label{btaudef}
 b(t)\equiv\frac{a(t)}{g(t)},\qquad d\tau\equiv\frac{dt}{f(t)},\qquad
 \bar\mu\equiv\frac{m}{2b(t)r}=\frac{m g(t)}{2a(t)r}.
\end{equation}
The consistent rainbow-parametrized metric is then
\begin{equation}\label{rainbowMetric}
 ds^2=-\frac{1}{f(t)^2}\left(\frac{1-\bar\mu}{1+\bar\mu}\right)^2dt^2
 +\frac{a(t)^2}{g(t)^2}(1+\bar\mu)^4\left(dr^2+r^2d\Omega^2\right).
\end{equation}
In the variables $(\tau,b)$, Eq.~\eqref{rainbowMetric} is exactly the standard McVittie metric,
\begin{equation}
 ds^2=-\bar\chi^2d\tau^2+b(\tau)^2(1+\bar\mu)^4\left(dr^2+r^2d\Omega^2\right),
 \qquad
 \bar\chi\equiv\frac{1-\bar\mu}{1+\bar\mu}.
\end{equation}
The areal radius and effective Hubble rate are
\begin{equation}\label{RbarH}
 \widetilde R=b r(1+\bar\mu)^2,
 \qquad
 \mathcal H\equiv\frac{1}{b}\frac{db}{d\tau}
 =f\left(H-\frac{\dot g}{g}\right).
\end{equation}
Here $\dot{\bar\mu}/\bar\mu=G_g-H$. Consequently, the $\bar\mu$ dependence of the spatial factor cancels the lapse factor in the extrinsic curvature:
\begin{equation}\label{consistentExtrinsic}
 K^i{}_j=-\mathcal H\,\delta^i{}_j,
 \qquad
 D_j\!\left(K^j{}_r-\delta^j{}_rK\right)=0.
\end{equation}
This cancellation distinguishes the consistent non-accreting construction from the direct ansatz. The factor $(1+\mu)/(1-\mu)$ belongs to Eq.~\eqref{directExtrinsic}; inserting it into $\mathcal H$ or the perfect-fluid Friedmann equations would restore an unwanted radial momentum density.
Accordingly, the later perfect-fluid Friedmann equations, trapping-horizon thermodynamics, and observational closure retain their stated form. The factor diagnoses the radial energy flux required by the direct ansatz rather than an additional homogeneous expansion term.
Consequently,
\begin{equation}
 \bar\chi^2=1-\frac{2m}{\widetilde R},
\end{equation}
and the central McVittie mass remains the constant $m$.

The transformation to $(\tau,b)$ is exact for every smooth positive pair $f(t),g(t)$. Hence all curvature scalars, trapping-horizon relations, and perfect-fluid Einstein equations for one fixed energy history depend only on $b(\tau)$ and $m$. The separate functions $f$ and $g$ cannot be recovered from those invariants. A physical rainbow effect requires additional structure that compares distinct probe energies or specifies how energy-dependent rods, clocks, couplings, or matter equations are related. This identifiability statement prevents a coordinate reparametrization from being counted as an independent gravitational correction.

\subsection{Rainbow-parametrized Friedmann equations}

Because Eq.~\eqref{rainbowMetric} is McVittie in $(\tau,b)$, the Einstein equations take the exact form
\begin{align}
 \mathcal H^2&=\frac{8\pi}{3}\rho,\label{rainF1tau}\\
 \mathcal H'&=-4\pi(\rho+p)\bar\chi,
 \qquad {}'\equiv\frac{d}{d\tau}.\label{rainF2tau}
\end{align}
Equivalently, the pressure profile is
\begin{equation}\label{rainPressure}
 8\pi p=-3\mathcal H^2-\frac{2\mathcal H'}{\bar\chi}.
\end{equation}
The normalized comoving four-velocity is
\begin{equation}
 u^\tau=\bar\chi^{-1},
 \qquad\text{or equivalently}\qquad
 u^t=\frac{f}{\bar\chi}.
\end{equation}
In terms of the original coordinate $t$, Eqs.~\eqref{rainF1tau}--\eqref{rainF2tau} become
\begin{equation}\label{rainF1t}
 \left(H-\frac{\dot g}{g}\right)^2=\frac{8\pi}{3f^2}\rho,
\end{equation}
and
\begin{equation}\label{rainF2t}
 \dot H-\frac{\ddot g}{g}+\left(\frac{\dot g}{g}\right)^2
 +\frac{\dot f}{f}\left(H-\frac{\dot g}{g}\right)
 =-\frac{4\pi}{f^2}(\rho+p)\sqrt{1-\frac{2m}{\widetilde R}}.
\end{equation}
The conservation equation is
\begin{equation}\label{rainCont}
 \dot\rho+3\left(H-\frac{\dot g}{g}\right)
 \sqrt{1-\frac{2m}{\widetilde R}}\,(\rho+p)=0.
\end{equation}
Differentiating Eq.~\eqref{rainF1t} and using Eq.~\eqref{rainCont} reproduces Eq.~\eqref{rainF2t}, providing a direct internal consistency check. In the limit $f=g=1$, Eqs.~\eqref{rainF1t}--\eqref{rainCont} reduce to Eqs.~\eqref{frclassic1}, \eqref{frclassic2}, and \eqref{continuityClass}.

Equations~\eqref{rainF1t}--\eqref{rainCont} are the standard McVittie equations expressed through the redundant variables $(t,a,f,g)$. They become physically predictive as rainbow equations after a microscopic prescription relates $f$ and $g$ to probe energy and supplies an operational rule for comparing energy sectors.

\section{Rainbow Apparent-Horizon Geometry}

In $(\tau,r)$ coordinates,
\begin{equation}
 \partial_\tau\widetilde R=\mathcal H\widetilde R\bar\chi,
 \qquad
 \partial_r\widetilde R=b(1-\bar\mu^2).
\end{equation}
The trapping-horizon condition is therefore
\begin{equation}\label{rainHorizon}
 1-\frac{2m}{\widetilde R_A}-\mathcal H^2\widetilde R_A^2=0,
\end{equation}
or, in terms of $t$,
\begin{equation}
 1-\frac{2m}{\widetilde R_A}
 -f^2\left(H-\frac{\dot g}{g}\right)^2\widetilde R_A^2=0.
\end{equation}
For an expanding sector $\mathcal H>0$, two positive horizons exist when $0<3\sqrt{3}\,m\mathcal H<1$. The black-hole branch lies in $2m<\widetilde R_{Ab}<3m$, the cosmological branch lies in $\widetilde R_{Ac}>3m$, and the equality $3\sqrt{3}\,m\mathcal H=1$ gives the degenerate radius $3m$.
The dynamical-horizon surface gravity follows immediately from the standard McVittie result with $H\mapsto\mathcal H$:
\begin{equation}\label{rainKappa}
 \kappa_A=\frac{1}{\widetilde R_A}\left(\frac{3m}{\widetilde R_A}-1\right)
 -\frac{\mathcal H'\widetilde R_A}{2\sqrt{1-2m/\widetilde R_A}}.
\end{equation}
The physical horizon temperature and entropy are
\begin{equation}
 T_A=\frac{|\kappa_A|}{2\pi},\qquad
 S_A=\frac{\mathcal A_A}{4}=\pi\widetilde R_A^2,
 \qquad \mathcal A_A=4\pi\widetilde R_A^2.
\end{equation}
The area law in this analysis uses a constant Newton coupling. An extension with $G(E)$ carries the corresponding energy dependence into the horizon entropy and thermodynamic equations.

\section{Thermodynamic Reconstruction and Branch Selection}

Energy-dependent horizon thermodynamics in Gravity's Rainbow has been developed for Schwarzschild black holes, general black objects, extended black branes, and thermodynamic phase structure \cite{galan2006,lingli2007,ali2014,alifaizal2015remnant,ashour2016,feng2017}. Dynamical-horizon thermodynamics and McVittie applications provide the complementary geometric framework for energy flux, tunneling, and phase behavior \cite{jiang2011,akbar2017,abdusattar2022}. This section implements the apparent-horizon unified-first-law formalism in spherical symmetry \cite{hayward1998}. The thermodynamic interpretation follows the equation-of-state perspective \cite{jacobson1995}, while the explicit variables $W$, $\Psi_a$, and the Misner--Sharp energy provide the trapping-horizon realization used in the calculation.

For a spherically symmetric spacetime, the Misner--Sharp energy is
\begin{equation}\label{MSrain}
 E=\frac{\widetilde R}{2}
 \left(1-h^{ab}\partial_a\widetilde R\partial_b\widetilde R\right)
 =m+\frac{1}{2}\mathcal H^2\widetilde R^3.
\end{equation}
When the first Friedmann equation holds, this becomes
\begin{equation}
 E=m+\frac{4\pi}{3}\rho\widetilde R^3.
\end{equation}
The work density and energy-supply one-form are \cite{hayward1998}
\begin{equation}
 W=-\frac12T^{ab}h_{ab}=\frac{\rho-p}{2},
\end{equation}
\begin{equation}
 \Psi_a=T_a{}^b\partial_b\widetilde R+W\partial_a\widetilde R.
\end{equation}
For the normalized comoving perfect fluid,
\begin{equation}\label{PsiComp}
 \Psi_\tau=-\frac{\rho+p}{2}\mathcal H\widetilde R\bar\chi,
 \qquad
 \Psi_r=\frac{\rho+p}{2}\,b(1-\bar\mu^2),
\end{equation}
so that
\begin{equation}\label{PsiForm}
 \Psi=\frac{\rho+p}{2}
 \left[d\widetilde R-2\mathcal H\widetilde R\bar\chi\,d\tau\right].
\end{equation}
The unified first law is
\begin{equation}\label{UFL}
 dE=\mathcal A\Psi+W\,dV,
 \qquad
 \mathcal A=4\pi\widetilde R^2,
 \qquad
 V=\frac{4\pi}{3}\widetilde R^3.
\end{equation}
On the apparent horizon, Eq.~\eqref{rainHorizon} implies $E_A=\widetilde R_A/2$. We write $p_A\equiv p(\tau,\widetilde R_A)$ for the pressure evaluated on the selected horizon branch.

Let
\begin{equation}
 \xi=\partial_\tau+\widetilde R_A'\partial_{\widetilde R}
\end{equation}
be tangent to the horizon trajectory in the $(\tau,\widetilde R)$ plane. Here Eq.~\eqref{PsiForm}, rather than the components in Eq.~\eqref{PsiComp}, supplies the one-form in the $(\tau,\widetilde R)$ chart. The matter heat flux through the horizon is the projected energy-supply term,
\begin{equation}\label{heatflux}
 \delta Q=\langle\mathcal A\Psi,\xi\rangle d\tau
 =2\pi\widetilde R_A^2(\rho+p_A)
 \left(\widetilde R_A'-2\mathcal H\widetilde R_A\bar\chi_A\right)d\tau.
\end{equation}
For the oriented trapping-horizon first law one uses the signed horizon temperature $T_{\rm s}=\kappa_A/(2\pi)$ \cite{hayward1998}; the sign can equivalently be placed in the definition of the outward heat flux. Since
\begin{equation}
 dS_A=2\pi\widetilde R_A\widetilde R_A' d\tau,
\end{equation}
the Clausius relation $\delta Q=T_{\rm s}dS_A$ is precisely the projection of Eq.~\eqref{UFL} along the trapping horizon.

The unified first law with the matter energy-supply form is equivalent to the spherically symmetric Einstein dynamics \cite{hayward1998}. The calculation below therefore reconstructs and checks the Friedmann equation and identifies the degeneracy of the horizon projection, while the field equations supply its dynamical foundation.

To see the dynamical content explicitly, differentiate Eq.~\eqref{rainHorizon}:
\begin{equation}\label{hordiff}
 \widetilde R_A'
 =\frac{\mathcal H\mathcal H'\widetilde R_A^4}{3m-\widetilde R_A},
\end{equation}
for a nondegenerate branch $\widetilde R_A\neq3m$. In the expanding sector, Eq.~\eqref{rainHorizon} gives $\bar\chi_A=\mathcal H\widetilde R_A$. Substitution of Eqs.~\eqref{rainKappa}, \eqref{heatflux}, and \eqref{hordiff} then gives the exact residual
\begin{align}\label{clausiusFactor}
 \frac{\delta Q-T_{\rm s}dS_A}{d\tau}
 ={}&\frac{\mathcal H\widetilde R_A^4}
 {2\bar\chi_A(3m-\widetilde R_A)}
 \left[\mathcal H'+4\pi\bar\chi_A(\rho+p_A)\right]\nonumber\\
 &\times\left[\mathcal H'\widetilde R_A^2
 -2\mathcal H(3m-\widetilde R_A)\right].
\end{align}
The nondegenerate dynamical factor gives
\begin{equation}\label{thermoF2}
 \mathcal H'=-4\pi(\rho+p_A)\bar\chi_A.
\end{equation}
This is the horizon evaluation of Eq.~\eqref{rainF2tau}. The remaining factor has a sharper interpretation. If
\begin{equation}\label{degenerateThermo}
 \mathcal H'\widetilde R_A^2
 =2\mathcal H(3m-\widetilde R_A),
\end{equation}
then Eqs.~\eqref{rainKappa}, \eqref{hordiff}, and $\bar\chi_A=\mathcal H\widetilde R_A$ give
\begin{equation}
 \kappa_A=0,
 \qquad
 \widetilde R_A'-2\mathcal H\widetilde R_A\bar\chi_A=0.
\end{equation}
Both sides of the projected Clausius relation vanish. Equation~\eqref{degenerateThermo} is therefore a zero-temperature, zero-heat-flux degeneracy that supplies no matter evolution law. In the $m\to0$ limit it takes the form $\mathcal H'=-2\mathcal H^2$. The thermodynamic reconstruction of the perfect-fluid McVittie dynamics is Eq.~\eqref{thermoF2}.

Finally, the conservation law in $\tau$ time is
\begin{equation}\label{contTau}
 \rho'+3\mathcal H\bar\chi(\rho+p)=0.
\end{equation}
Evaluating Eq.~\eqref{contTau} on the same horizon branch and combining it with Eq.~\eqref{thermoF2} gives
\begin{equation}
 \frac{d}{d\tau}\left(\mathcal H^2-\frac{8\pi}{3}\rho\right)=0.
\end{equation}
Hence
\begin{equation}
 \mathcal H^2=\frac{8\pi}{3}\rho+C,
\end{equation}
where $C$ is an integration constant. Spatial flatness by itself leaves this additive constant free. If $\rho$ excludes vacuum energy, one identifies $C=\Lambda/3$; if $\rho$ denotes the total density, the redefinition $\rho_{\rm tot}=\rho+3C/(8\pi)$ with $p_{\rm vac}=-3C/(8\pi)$ sets $C=0$ and recovers Eq.~\eqref{rainF1tau}. The horizon Clausius relation, local conservation evaluated on the horizon, and the McVittie trapping-horizon geometry thereby reconstruct the Friedmann dynamics under the stated branch and matter assumptions.

\section{Phenomenological Illustration and Low-Energy Constraints}

Equations~\eqref{rainF1tau}--\eqref{rainCont} reduce exactly to standard McVittie dynamics for each prescribed energy history and therefore leave the late-time correction to $H(z)$ unspecified. Rainbow cosmology has connected energy-dependent metrics with modified Friedmann evolution, nonsingular solutions, primordial fluctuations, Gauss--Bonnet dynamics, and loop-quantum-cosmology effective equations \cite{ling2007,awad2013,amelino2013,hendi2016gb,gorji2017}. Motivated by that literature, we now define a separate one-parameter observational closure. The likelihood below tests this closure itself. Translation into a fundamental rainbow model requires a microscopic choice of $f(E/E_{\rm P})$, $g(E/E_{\rm P})$, energy evolution, and matter propagation.

\subsection{Minimal observational closure}

Let $H_{\rm op}(z)$ denote the expansion rate assigned by a specified operational protocol and define $\Xi(z)\equiv H_{\rm op}(z)/H_{\Lambda\mathrm{CDM}}(z)$. We normalize the response at the present epoch, $\Xi(0)=1$. If the response is analytic around $z=0$, its leading correction is
\begin{equation}\label{XiExpansion}
 \Xi(z)=1+\epsilon z+\mathcal O(z^2),
\end{equation}
where $\epsilon=\Xi'(0)$. A model that also adopts the standard probe-redshift law $E(z)=E_0(1+z)$ can relate $\epsilon$ to the derivative of its microscopic rainbow response at $E_0$. Choosing flat $\Lambda$CDM as the reference background, with vacuum energy included in the total density, gives the benchmark closure
\begin{equation}\label{HobsRG}
 H_{\rm op}(z)\equiv H_{\rm RG}(z)
 =H_0\sqrt{\Omega_m(1+z)^3+1-\Omega_m}\,(1+\epsilon z).
\end{equation}
Equation~\eqref{HobsRG} uses the linear term as the fitted closure over the data range and restricts the parameter domain to $1+\epsilon z>0$. It is nested: $\epsilon=0$ reproduces flat $\Lambda$CDM exactly. Without an additional microscopic map, $\epsilon$ measures an effective redshift response and leaves $f$ and $g$ individually unidentified.

\subsection{Leading microscopic map for analytic rainbow functions}

The phenomenological coefficient can be connected to a specified low-energy sector under additional assumptions. Write $y\equiv E/E_{\rm P}$ and suppose the first nonzero corrections have the common order $n\geq1$,
\begin{equation}\label{microFunctions}
 f(y)=1+\alpha_n y^n+\mathcal O(y^{n+1}),\qquad
 g(y)=1+\beta_n y^n+\mathcal O(y^{n+1}).
\end{equation}
If the reference energy obeys the standard redshift law $\dot y/y=-H$, Eq.~\eqref{RbarH} gives
\begin{equation}\label{microHmap}
 \frac{\mathcal H}{H}
 =1+(\alpha_n+n\beta_n)y^n+\mathcal O(y^{n+1}).
\end{equation}
Conditionally identifying the operational response with the present-normalized ratio in Eq.~\eqref{microHmap}, and using $y(z)=y_0(1+z)$, yields
\begin{equation}\label{microResponse}
 \Xi_n(z)=1+A_n y_0^n\left[(1+z)^n-1\right]
 +\mathcal O(y_0^{n+1}),
 \qquad A_n\equiv\alpha_n+n\beta_n.
\end{equation}
Consequently, the local slope in Eq.~\eqref{XiExpansion} is
\begin{equation}\label{epsilonMap}
 \boxed{\epsilon=n(\alpha_n+n\beta_n)
 \left(\frac{E_0}{E_{\rm P}}\right)^n}.
\end{equation}
For $n=1$, the fitted closure is the leading response itself. For $n>1$, it is only the tangent at $z=0$, and the curvature in Eq.~\eqref{microResponse} becomes relevant over the full supernova range. Equation~\eqref{epsilonMap} also makes the identifiability limit quantitative: one measured slope constrains the combination $\alpha_n+n\beta_n$ only after $n$ and $E_0$ are specified; it does not determine the two rainbow functions separately.

To first order, the observable departure is transparent:
\begin{equation}\label{deltaH}
 \frac{\Delta H}{H_{\Lambda\mathrm{CDM}}}=\epsilon z.
\end{equation}
The corresponding deceleration parameter is
\begin{equation}\label{qRG}
 q_{\rm RG}(z)=q_{\Lambda\mathrm{CDM}}(z)
 +\frac{(1+z)\epsilon}{1+\epsilon z},
\end{equation}
so that the same parameter that changes the expansion history also changes the acceleration-transition redshift.  This provides an immediately testable consistency relation between $H(z)$ and distance measurements.

\subsection{Operational assumptions}

The joint likelihood implements four assumptions. First, the catalog redshift retains its standard operational meaning. Second, photon propagation follows the usual radial-distance integral and Etherington distance duality \cite{etherington1933}, with the closure acting through $H_{\rm op}(z)$. Third, one common effective $\epsilon$ describes the cosmic-chronometer and supernova sectors. Fourth, the local McVittie mass affects horizon and pressure observables while remaining outside the homogeneous background fit. These assumptions define a controlled one-parameter test. A chromatic rainbow model can instead assign different responses to photon bands and clock constructions and must derive its redshift, null-geodesic, and luminosity-distance relations before applying the numerical bound below.

\subsection{Cosmic-chronometer likelihood}

We use the 32 cosmic-chronometer measurements compiled in \cite{favale2023}, spanning $0.07<z<1.965$. For the subset with publicly characterized correlated systematics, the initial-mass-function and stellar-population-synthesis contributions are reconstructed with the covariance prescription in \cite{moresco2020}, while the published diagonal uncertainties of the 32-point compilation are retained. This produces the quadratic likelihood
\begin{equation}\label{chiCC}
 \chi^2_{\rm CC}=\Delta\mathbf H^{\rm T}C_{\rm CC}^{-1}\Delta\mathbf H,
 \qquad
 \Delta H_i=H_i^{\rm obs}-H_{\rm RG}(z_i).
\end{equation}
The chronometer covariance model combines the published diagonal uncertainties with the public correlated initial-mass-function and stellar-population-synthesis contributions characterized in \cite{moresco2020}. A diagonal-error cross-check shifts the fitted rainbow amplitude by approximately $10^{-3}$ in the joint analysis, demonstrating numerical stability against this covariance choice.

The cosmic-chronometer fit gives $H_0=71.87~\mathrm{km\,s^{-1}\,Mpc^{-1}}$, $\Omega_m=0.2743$, and $\chi^2=12.290$ for flat $\Lambda$CDM. The rainbow closure gives $H_0=71.81~\mathrm{km\,s^{-1}\,Mpc^{-1}}$, $\Omega_m=0.2824$, $\epsilon=-0.0078$, and $\chi^2=12.288$. The statistic changes by $\Delta\chi^2\simeq0.001$, while $\Delta\mathrm{AIC}\simeq+2.00$ and $\Delta\mathrm{BIC}\simeq+3.46$ select the nested general-relativistic limit.

The corresponding reduced statistics are $12.290/30=0.410$ and $12.288/29=0.424$. Their small values indicate conservative chronometer errors relative to the observed scatter. Parameter discrimination from this subset is consequently weak, while the nested-model information-criterion differences remain arithmetically well defined.

\begin{figure}[htbp]
 \centering
 \includegraphics[width=0.88\textwidth]{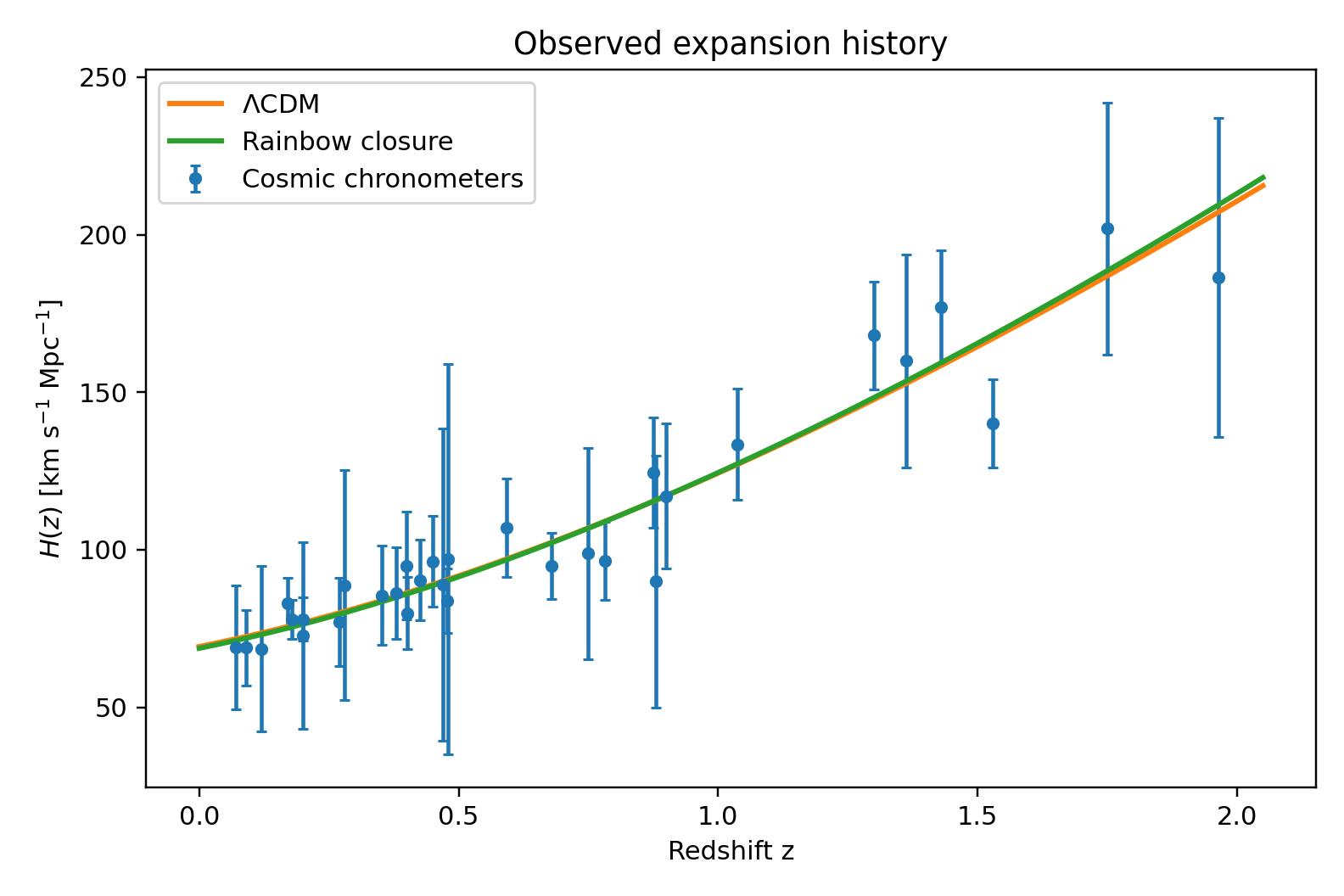}
 \caption{Cosmic-chronometer expansion history and the best fits obtained after combining the chronometers with the Pantheon+ shape likelihood and the Planck matter-density prior.  The close overlap of the two curves visualizes the percent-level fitted deformation.}
 \label{fig:CCfit}
\end{figure}

\begin{figure}[htbp]
 \centering
 \includegraphics[width=0.88\textwidth]{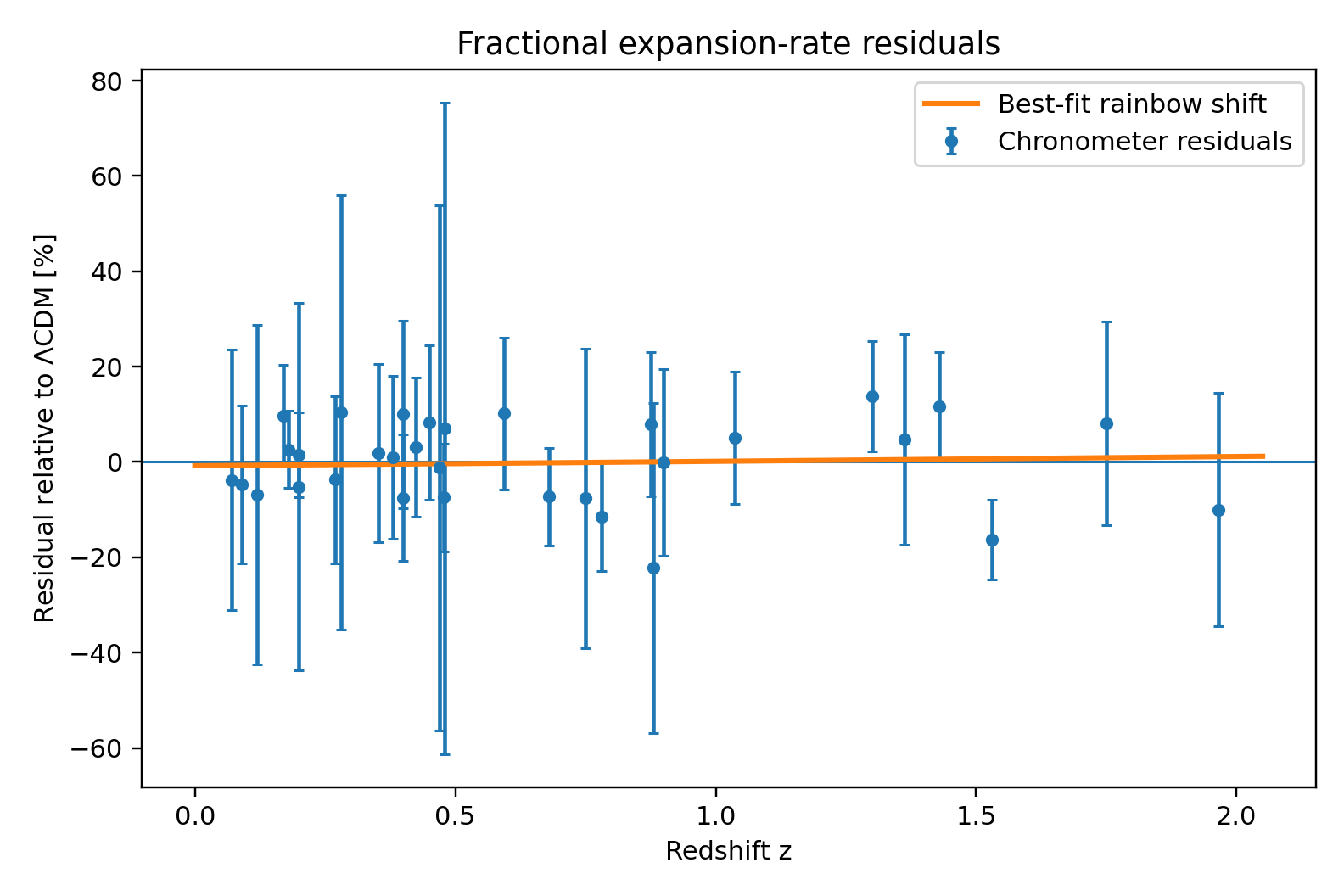}
 \caption{Fractional chronometer residuals relative to the joint best-fit $\Lambda$CDM model.  The solid model curve is the shift produced by the best-fit rainbow closure.  Its magnitude remains at the percent level over the measured redshift range.}
 \label{fig:CCres}
\end{figure}

\subsection{Pantheon+ with the released full covariance}

\begin{figure}[htbp]
 \centering
 \includegraphics[width=0.88\textwidth]{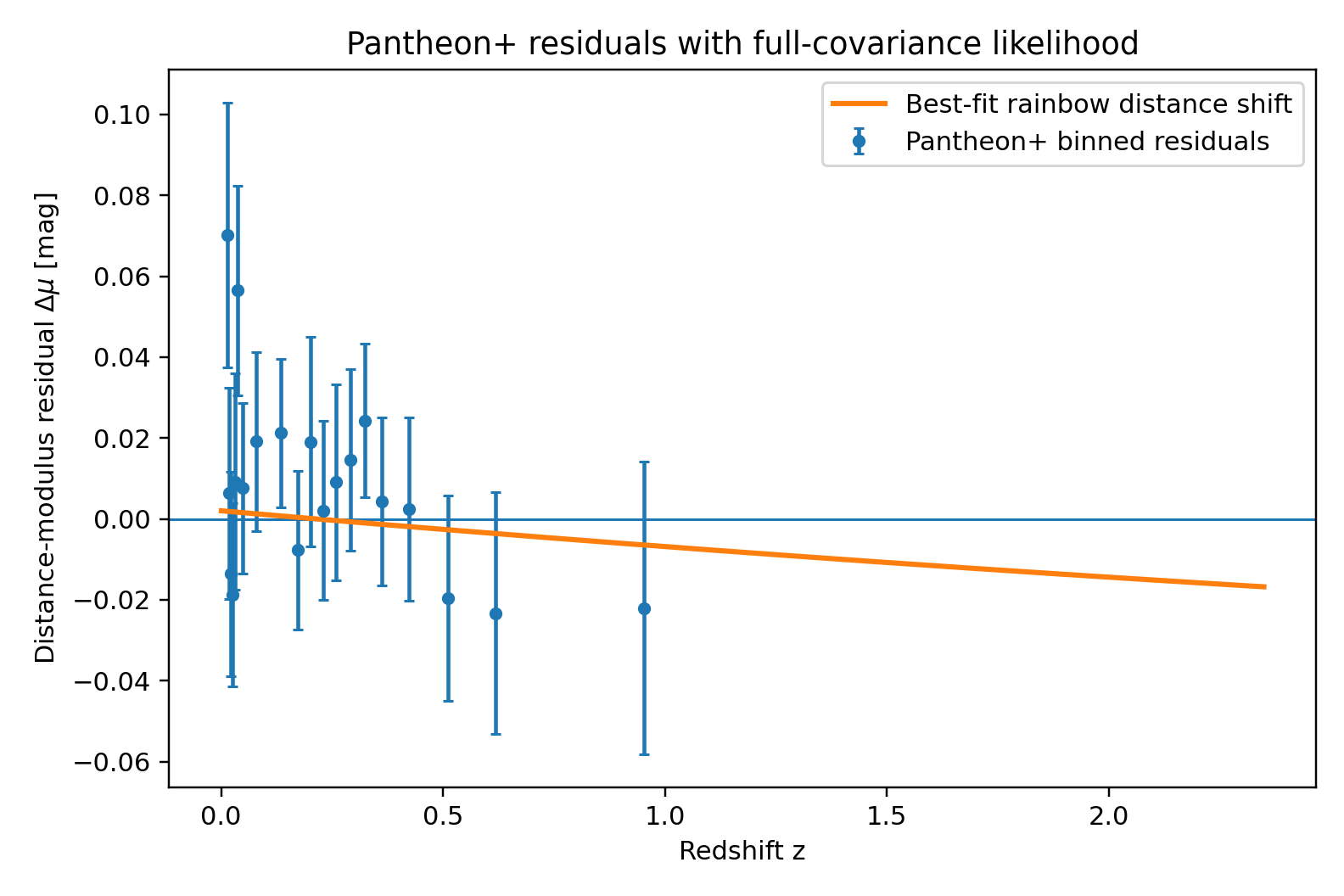}
 \caption{Pantheon+ residuals after linear binning for visualization. The likelihood uses the complete released statistical-plus-systematic covariance of the 1580 selected light curves, while the displayed error bars provide the binned visual summary. The curve shows the distance-modulus shift between the best-fit rainbow closure and the best-fit $\Lambda$CDM model.}
 \label{fig:SNres}
\end{figure}

We additionally use the Pantheon+ data release \cite{scolnic2022,brout2022}. Applying the catalog mask \texttt{IS\_CALIBRATOR}$=0$ and $z_{\rm HD}>0.01$ leaves 1580 light curves. The released $1701\times1701$ statistical-plus-systematic covariance matrix is reduced with the same Boolean mask on both axes. Under the operational propagation assumptions stated above, the dimensionless luminosity-distance shape for the flat model is
\begin{equation}\label{DLobs}
 \mathcal D_L(z_{\rm hel},z_{\rm HD})=(1+z_{\rm hel})
 \int_0^{z_{\rm HD}}\frac{dz'}{E_{\rm RG}(z')},
 \qquad E_{\rm RG}\equiv\frac{H_{\rm RG}}{H_0}.
\end{equation}
An additive magnitude zero point, which is exactly degenerate with the absolute distance scale in a shape-only supernova analysis, is marginalized analytically.  If $\Delta\boldsymbol\mu$ denotes the residual before this marginalization and $\mathbf 1$ is a vector of ones, the effective statistic is
\begin{equation}\label{chiSN}
 \chi^2_{\rm SN}=
 \Delta\boldsymbol\mu^{\rm T}C_{\rm SN}^{-1}\Delta\boldsymbol\mu
 -\frac{\left(\mathbf 1^{\rm T}C_{\rm SN}^{-1}\Delta\boldsymbol\mu\right)^2}
 {\mathbf 1^{\rm T}C_{\rm SN}^{-1}\mathbf 1}.
\end{equation}
Terms independent of the cosmological parameters are omitted from Eq.~\eqref{chiSN}. After zero-point marginalization, the chronometer likelihood and external prior supply the $H_0$ information. The supernova likelihood uses the full covariance, while Fig.~\ref{fig:SNres} presents a binned visual compression of the residuals.

Using the low-redshift CC+Pantheon+ data, the three-parameter fit gives approximately
\begin{equation}\label{lowzfit}
 H_0=69.99\pm2.82~\mathrm{km\,s^{-1}\,Mpc^{-1}},\qquad
 \Omega_m=0.399\pm0.063,\qquad
 \epsilon=-0.083\pm0.067,
\end{equation}
where the quoted errors are local Gaussian estimates.  Relative to the nested $\Lambda$CDM fit, the low-redshift closure changes the minimum statistic by $\Delta\chi^2\simeq1.05$, corresponding to about $1.0$--$1.3\sigma$ depending on the projection. Equation~\eqref{lowzfit} also quantifies a substantial $\Omega_m$--$\epsilon$ degeneracy, with the matter-density shift tracking part of the redshift-dependent rainbow response.

\subsection{Joint constraint and the general-relativistic limit}

The joint analysis combines the chronometer and Pantheon+ likelihoods with the Planck 2018 base-$\Lambda$CDM matter-density constraint, $\Omega_m=0.315\pm0.007$ \cite{planck2018}, and leaves $H_0$ free. This Gaussian prior comes from the base-$\Lambda$CDM early-universe analysis rather than a CMB reanalysis of Eq.~\eqref{HobsRG}. The resulting constraint is therefore conditional on a late-time response that preserves the Planck inference of $\Omega_m$. The total statistic is
\begin{equation}\label{chiTotal}
 \chi^2_{\rm tot}=\chi^2_{\rm CC}+\chi^2_{\rm SN}
 +\left(\frac{\Omega_m-0.315}{0.007}\right)^2.
\end{equation}
The resulting best fit is
\begin{equation}\label{jointfit}
 H_0=68.75\pm2.63~\mathrm{km\,s^{-1}\,Mpc^{-1}},\qquad
 \Omega_m=0.3157\pm0.0070,
\end{equation}
with
\begin{equation}\label{epsfit}
 \epsilon=0.0106^{+0.0234}_{-0.0231}\quad(68\%~\mathrm{profile}),
 \qquad
 -0.0355\lesssim\epsilon\lesssim0.0585\quad(\Delta\chi^2\le4).
\end{equation}
The best-fit rainbow closure lowers the minimum statistic by $\Delta\chi^2=0.204$ relative to the nested $\epsilon=0$ model. The corresponding information-criterion shifts are $\Delta\mathrm{AIC}=+1.80$ and approximately $\Delta\mathrm{BIC}=+7.18$. The latter uses $N=1612$ low-redshift entries and treats the Gaussian prior separately from that count. Because the supernova data are strongly correlated and the matter constraint is an external prior, BIC supplies a heuristic summary here. Both criteria favor the nested $\epsilon=0$ closure.

\begin{figure}[htbp]
 \centering
 \includegraphics[width=0.88\textwidth]{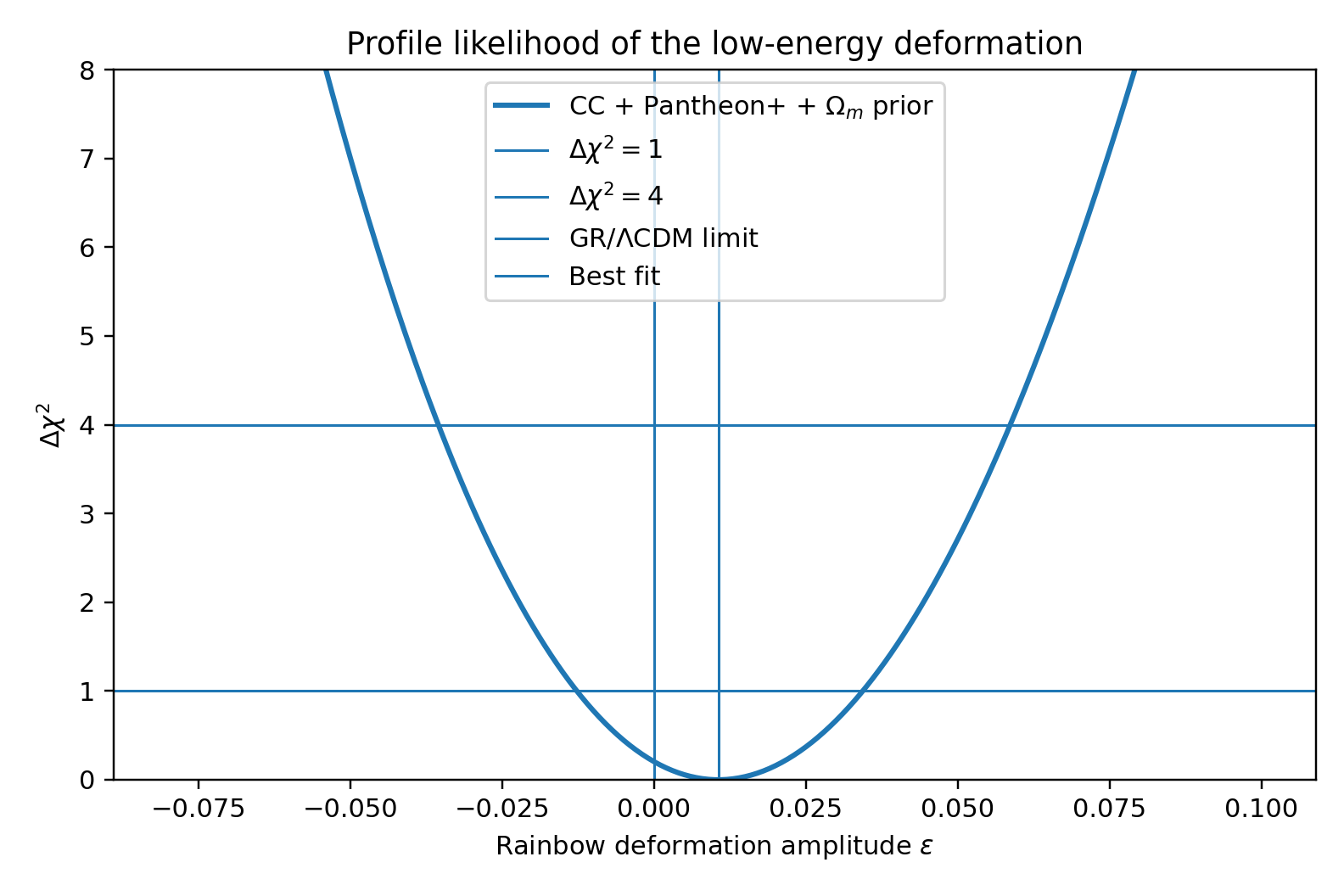}
 \caption{Profile likelihood of the leading low-energy rainbow amplitude after combining cosmic chronometers, the full-covariance Pantheon+ shape likelihood, and the Planck matter-density prior.  The profile places the general-relativistic value $\epsilon=0$ inside the one-standard-deviation region.}
 \label{fig:epsprofile}
\end{figure}

\begin{figure}[htbp]
 \centering
 \includegraphics[width=0.88\textwidth]{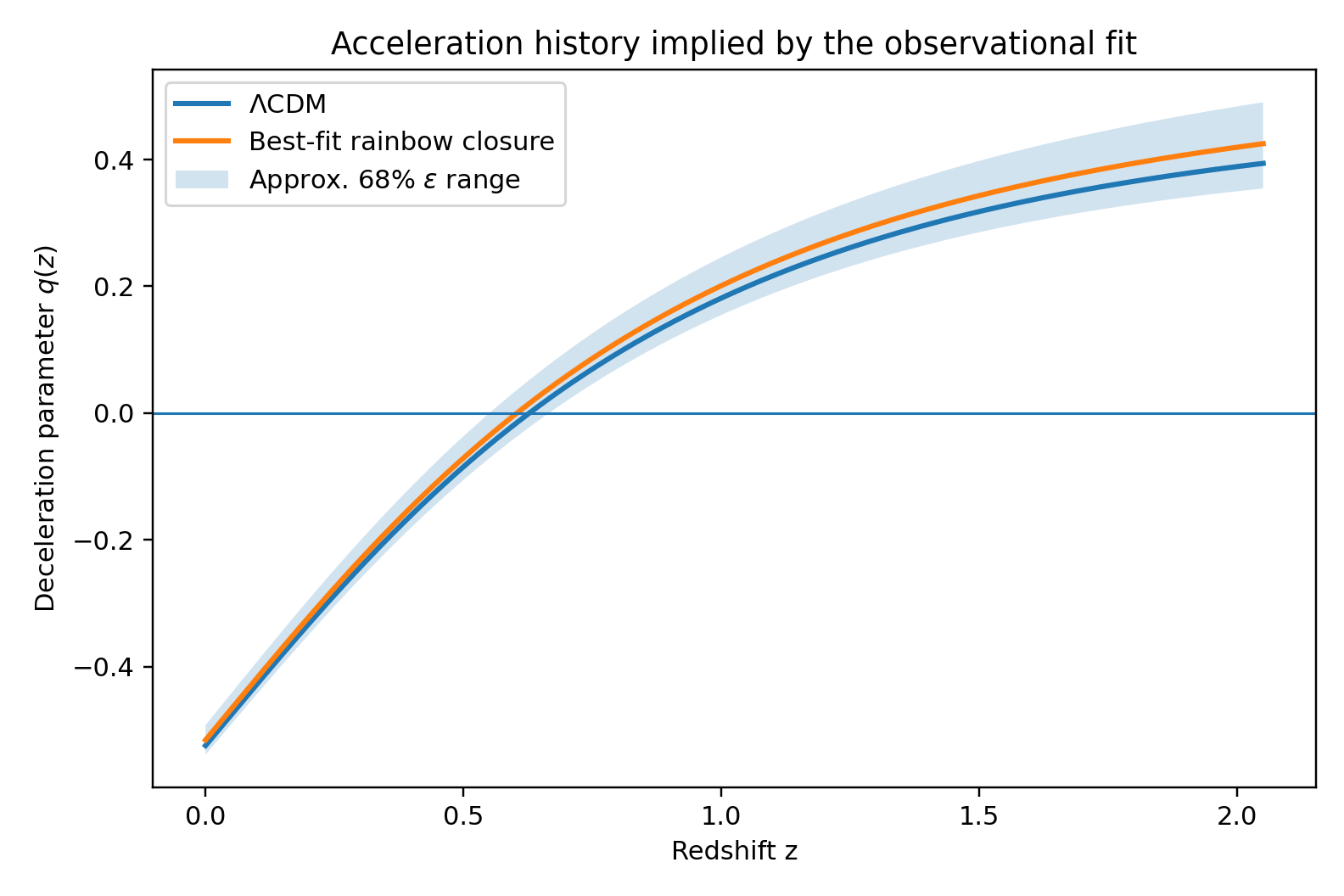}
 \caption{Acceleration history inferred from the observational closure.  The shaded region propagates the approximate 68\% profile interval of $\epsilon$ at the joint best-fit matter density.  The fitted profile corresponds to a modest displacement of the transition from deceleration to acceleration.}
 \label{fig:qhistory}
\end{figure}

For the best-fit parameters, Eq.~\eqref{qRG} gives $q_0\simeq-0.516$ and an acceleration-transition redshift $z_t\simeq0.604$.  The corresponding best-fit $\Lambda$CDM values are $q_0\simeq-0.525$ and $z_t\simeq0.628$.  Figure~\ref{fig:qhistory} places this transition shift within the profile range allowed by the fitted deformation amplitude.

\subsection{Observational scope of the constraints}

The observational likelihood constrains the effective background closure through $H(z)$ and luminosity-distance evolution. Equation~\eqref{rainF1tau} places the homogeneous expansion directly in $\mathcal H$ and $\rho$, while the McVittie central mass $m$ enters the pressure profile, trapping-horizon relation, surface gravity, and local geometry. This separation identifies the present fit as a background-expansion constraint and locates potential local McVittie signatures in observables such as horizon evolution, lensing, time delays, turnaround scales, and strong-field propagation.

The one-parameter fit assigns a common effective $\epsilon$ to the chronometer and supernova sectors. Within the stated assumptions, $\epsilon$ represents their shared redshift response. The nested value $\epsilon=0$ provides the exact $\Lambda$CDM reference point, and the fitted profile includes it within one standard deviation. The best fit is statistically aligned with zero and constrains the defined closure. The functions $f(E/E_{\rm P})$, $g(E/E_{\rm P})$, an energy-dependent Newton coupling, and chromatic propagation remain separate model inputs.

\subsection{Data and numerical reproducibility}

The Pantheon+ data vector and covariance are publicly released \cite{scolnic2022,brout2022}, and the chronometer compilation and covariance prescription are specified in Refs.~\cite{favale2023,moresco2020}. The accompanying source package records the numerical best fits, all ten final figures, a deterministic generator for the five dimensionless diagnostics, an independent verification script for the horizon, Clausius, model-selection, acceleration, and graphical-response identities, pinned figure dependencies, and a one-command build target. Independent end-to-end reproduction of the quoted confidence profile additionally requires the precise chronometer covariance matrix, likelihood implementation, optimizer settings, and profiling grid. Archival release of those numerical inputs alongside the manuscript will convert the present result record into a fully executable likelihood analysis.

\section{Dimensionless Graphical Diagnostics}

\begin{figure}[htbp]
 \centering
 \includegraphics[width=0.94\textwidth]{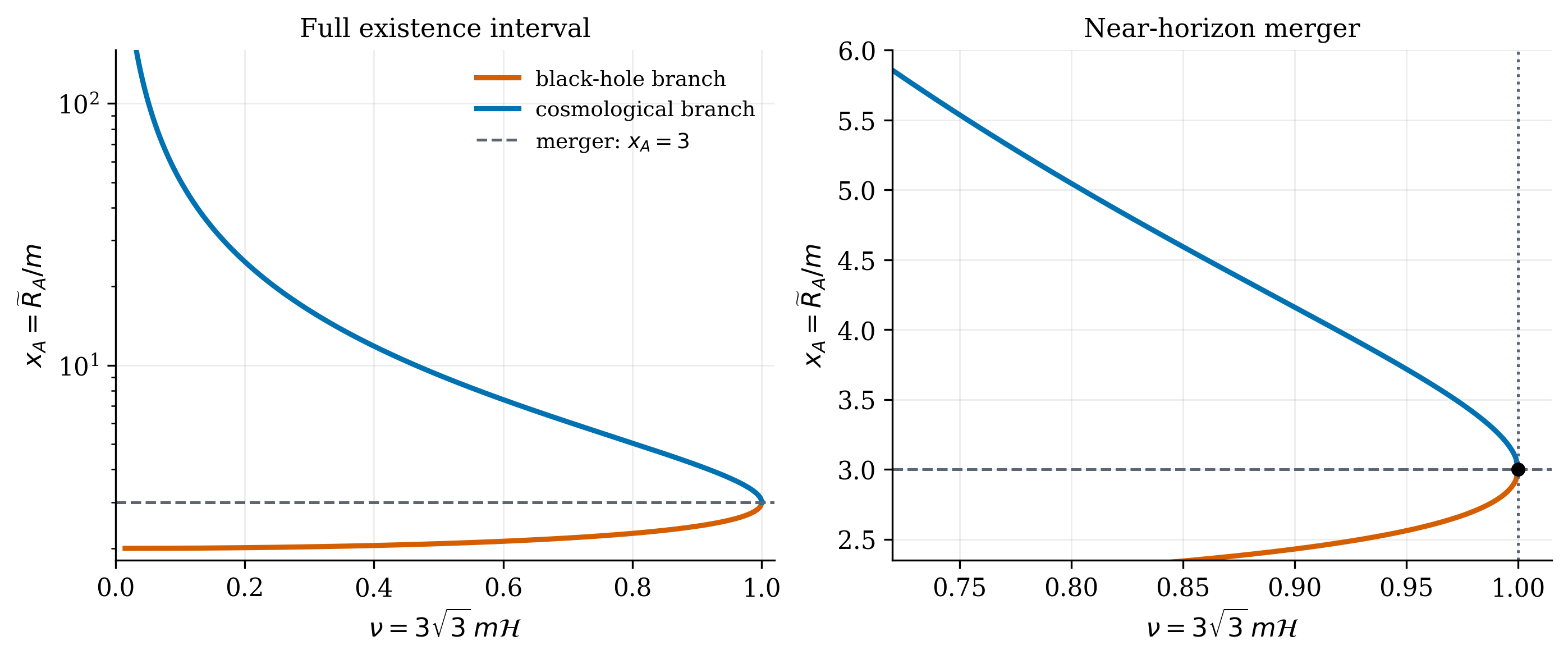}
 \caption{Exact positive roots of the dimensionless McVittie horizon cubic. The right panel magnifies the approach to the double-root merger at $\nu=1$ and $x_A=3$. No two-horizon exterior exists for $\nu>1$.}
 \label{fig:horizonBranches}
\end{figure}

The principal geometric and phenomenological results admit scale-free representations that expose their domains and limiting behavior. Define
\begin{equation}\label{dimensionlessHorizon}
 u\equiv m\mathcal H,\qquad
 \nu\equiv3\sqrt{3}\,u,\qquad
 x_A\equiv\frac{\widetilde R_A}{m}.
\end{equation}
The horizon equation becomes $u^2x_A^3-x_A+2=0$. For $0<\nu\leq1$, let $\theta=\frac13\sin^{-1}\nu$. Its two positive roots are
\begin{equation}\label{horizonRoots}
 x_{Ab}=\frac{2}{\sqrt3\,u}\sin\theta,
 \qquad
 x_{Ac}=\frac{2}{\sqrt3\,u}
 \sin\left(\theta+\frac{2\pi}{3}\right).
\end{equation}
Figure~\ref{fig:horizonBranches} shows the disjoint intervals $2<x_{Ab}<3<x_{Ac}$ and their common endpoint. The logarithmic panel also resolves the small-$u$ asymptotes $x_{Ab}\to2$ and $x_{Ac}\sim u^{-1}$.

\begin{figure}[htbp]
 \centering
 \includegraphics[width=0.78\textwidth]{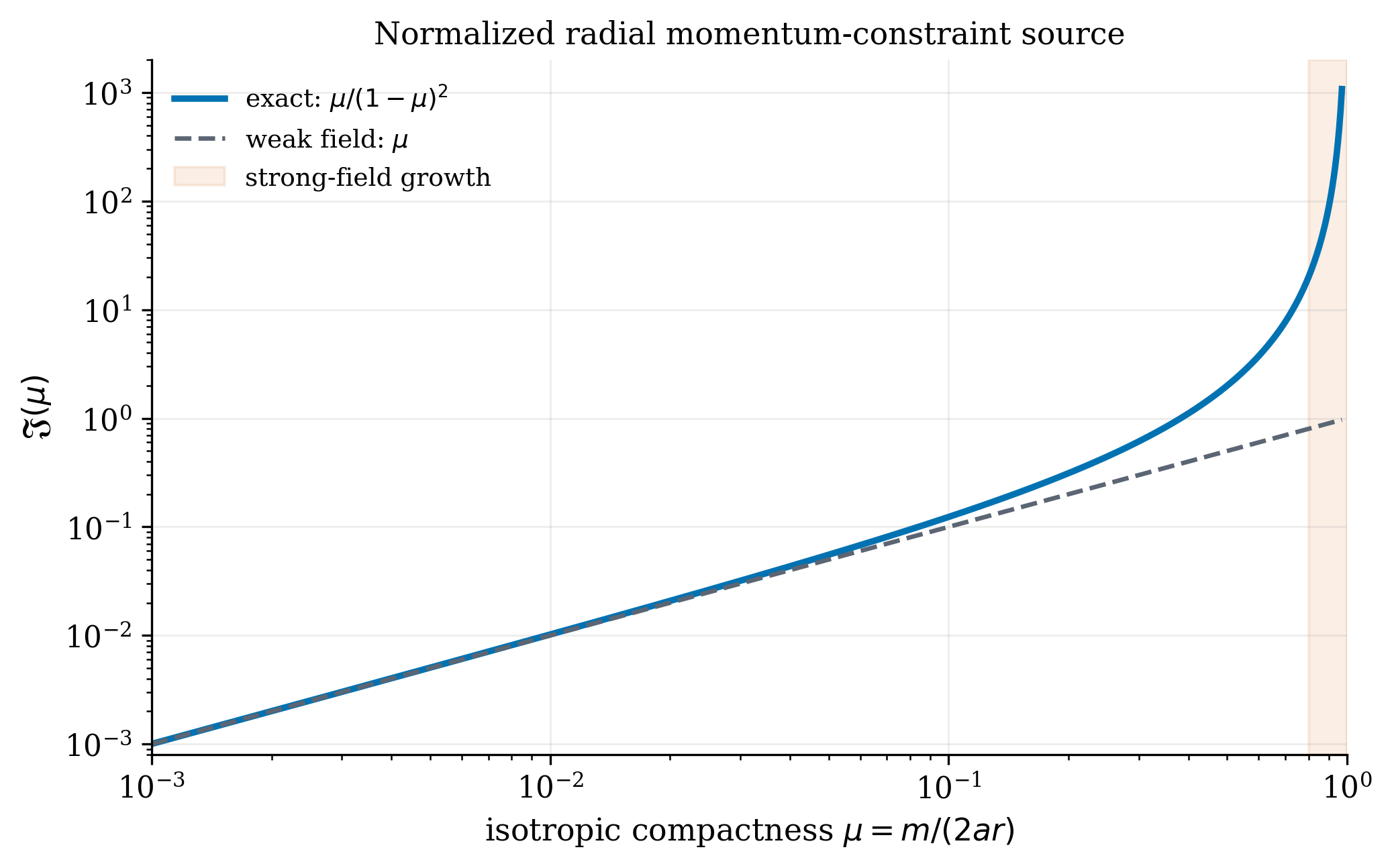}
 \caption{Scale-free radial momentum-constraint source generated by the factor $(1+\mu)/(1-\mu)$ in the direct rainbow ansatz. The exact curve follows the linear weak-field approximation only for $\mu\ll1$ and grows rapidly near the isotropic-coordinate boundary.}
 \label{fig:momentumProfile}
\end{figure}

The inconsistency of the direct ansatz can likewise be displayed without choosing $m$, $f$, or $\dot g$. Equation~\eqref{directExtrinsic} contains the factor $(1+\mu)/(1-\mu)$ multiplying $G_g$; the following profile displays the radial derivative generated by that factor. For $fG_g\neq0$, normalize Eq.~\eqref{naiveMomentum} as
\begin{equation}\label{normalizedMomentum}
 \mathfrak J(\mu)\equiv
 \frac{r}{4fG_g}D_j\!\left(K^j{}_r-\delta^j{}_rK\right)
 =\frac{\mu}{(1-\mu)^2}.
\end{equation}
Figure~\ref{fig:momentumProfile} confirms the weak-field behavior $\mathfrak J\sim\mu$ and the quadratic divergence as the isotropic boundary $\mu\to1^{-}$ is approached. Thus the radial momentum source is not a small uniform correction in the strong-field part of the chart.

\begin{figure}[htbp]
 \centering
 \includegraphics[width=0.94\textwidth]{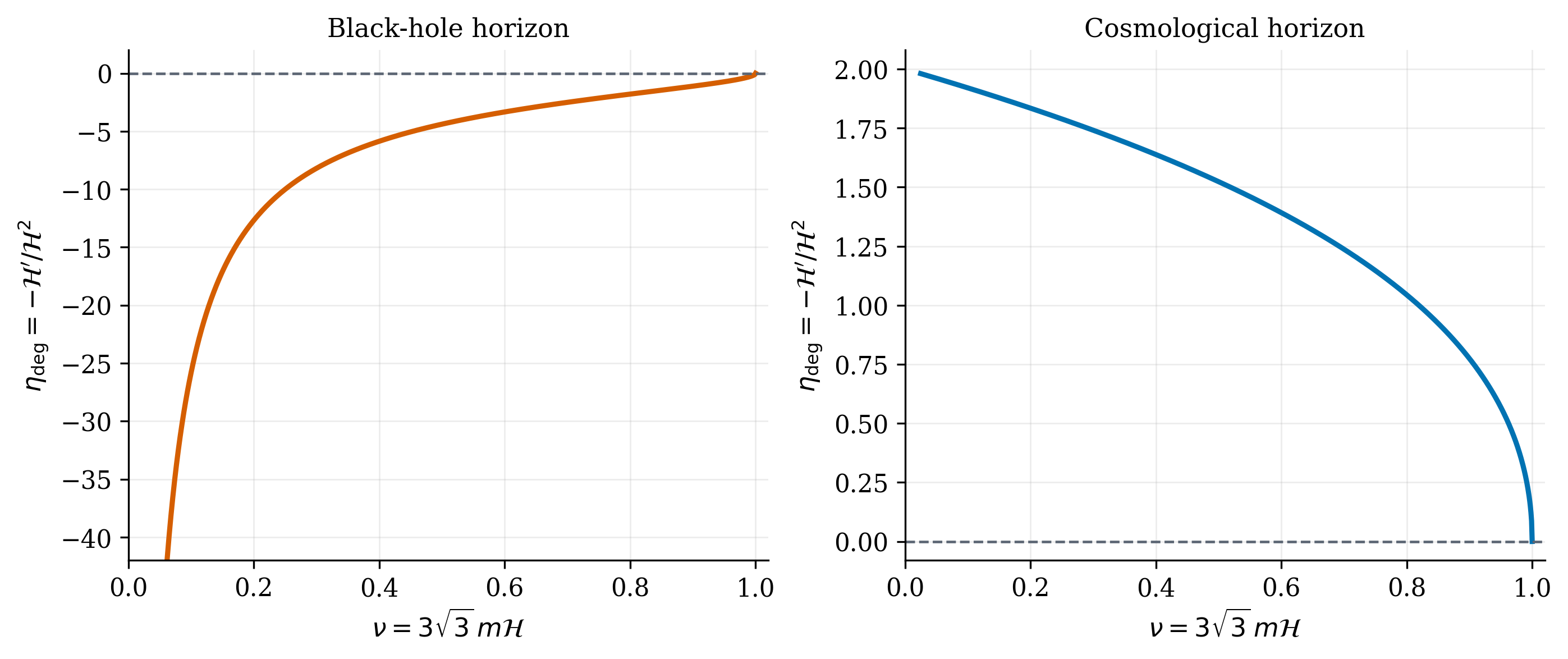}
 \caption{Zero-temperature, zero-flux Clausius degeneracy expressed through $\eta=-\mathcal H'/\mathcal H^2$. The distinct signs and limiting values on the two horizons confirm that this factor is a branch-dependent thermodynamic degeneracy rather than a universal matter evolution equation.}
 \label{fig:zeroTemperature}
\end{figure}

The thermodynamic degeneracy also has a compact phase-space form. Introduce the effective slow-roll parameter $\eta\equiv-\mathcal H'/\mathcal H^2$. On either horizon, Eq.~\eqref{rainKappa} reduces to
\begin{equation}\label{kappaDimensionless}
 m\kappa_A=\frac{3-x_A}{x_A^2}+\frac{\eta u}{2}.
\end{equation}
The zero-temperature factor in Eq.~\eqref{degenerateThermo} is therefore the curve
\begin{equation}\label{etaDegenerate}
 \eta_{\rm deg}(u)=\frac{2(x_A-3)}{u x_A^2}.
\end{equation}
Figure~\ref{fig:zeroTemperature} displays this locus on both roots. The black-hole degeneracy is negative and diverges as $u\to0$, while the cosmological degeneracy approaches $\eta_{\rm deg}\to2$ in that limit; both meet at zero when the horizons merge. This branch-sensitive structure is hidden if the second Clausius factor is reported without its geometric domain.

\begin{figure}[htbp]
 \centering
 \includegraphics[width=0.80\textwidth]{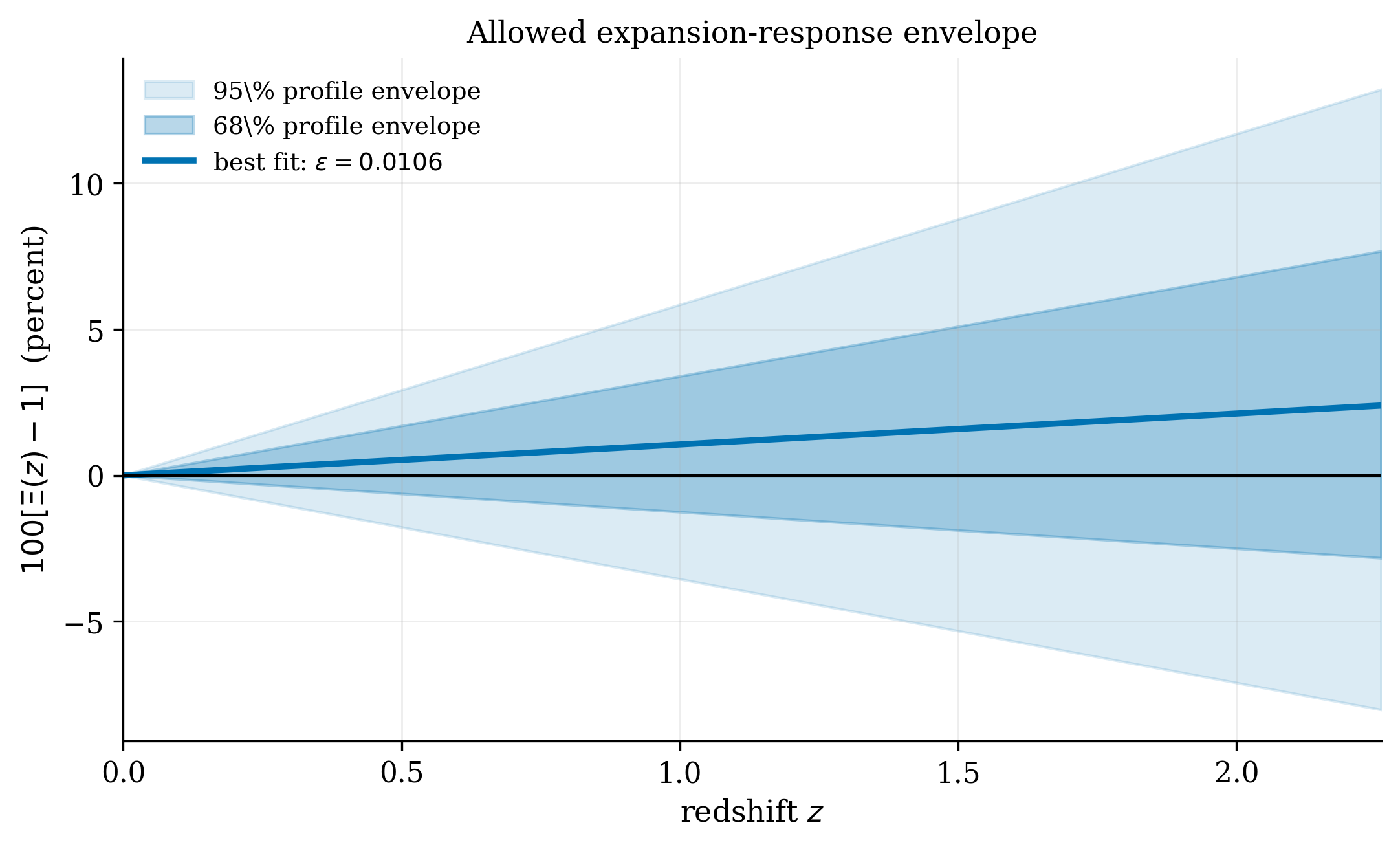}
 \caption{Fractional expansion response implied by the best-fit, 68\% profile, and $\Delta\chi^2\leq4$ intervals for $\epsilon$. The general-relativistic response remains inside the inner envelope over the plotted range.}
 \label{fig:responseEnvelope}
\end{figure}

Figure~\ref{fig:responseEnvelope} translates the profile intervals in Eq.~\eqref{epsfit} directly into the fractional expansion response. At the upper plotting redshift $z=2.26$, the best-fit shift is $2.39\%$, the approximate $68\%$ envelope is $[-2.83\%,7.68\%]$, and the $\Delta\chi^2\leq4$ envelope is $[-8.02\%,13.22\%]$. The widening is the deterministic consequence of the linear closure, not an additional uncertainty model.

\begin{figure}[htbp]
 \centering
 \includegraphics[width=0.80\textwidth]{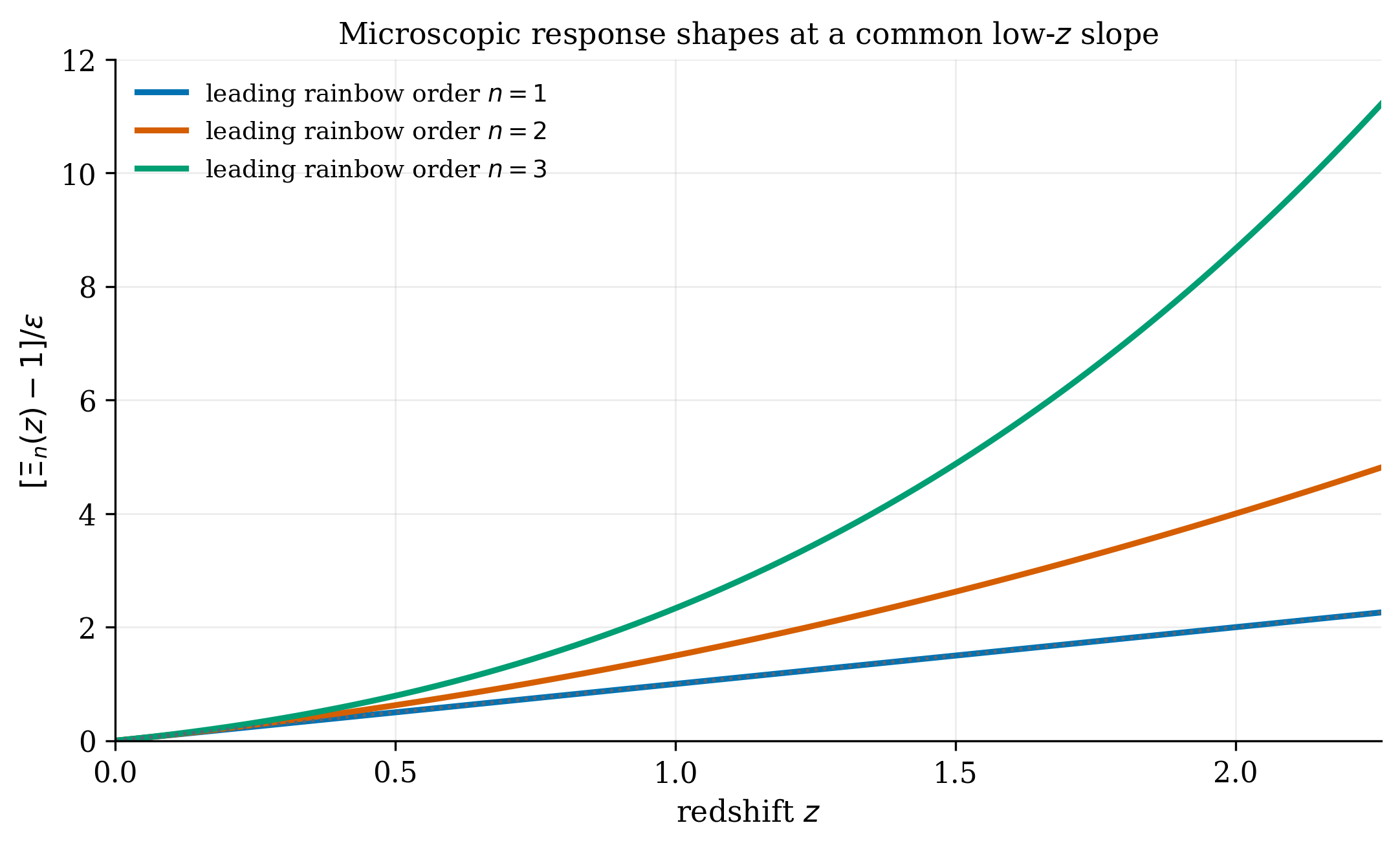}
 \caption{Leading microscopic response shapes normalized to the same derivative at $z=0$. Their separation at moderate redshift visualizes the model dependence that is lost in a one-parameter local Taylor slope.}
 \label{fig:microShapes}
\end{figure}

Finally, Eq.~\eqref{microResponse} predicts a family of shapes even when different microscopic orders share the same local slope. To leading order,
\begin{equation}\label{normalizedMicroResponse}
 \frac{\Xi_n(z)-1}{\epsilon}
 =\frac{(1+z)^n-1}{n}.
\end{equation}
Figure~\ref{fig:microShapes} shows that the $n=1$ closure is exactly linear at this order, whereas $n=2$ and $n=3$ develop appreciable curvature. The current single-slope fit can therefore be translated into Eq.~\eqref{epsilonMap} only after the leading microscopic order is fixed, and a future data analysis of $n>1$ models should fit Eq.~\eqref{microResponse} rather than extrapolating Eq.~\eqref{HobsRG} across the entire redshift interval.

All curves in Figs.~\ref{fig:horizonBranches}--\ref{fig:microShapes} are generated by the supplied \nolinkurl{generate_additional_figures.py} script directly from Eqs.~\eqref{horizonRoots}, \eqref{normalizedMomentum}, \eqref{etaDegenerate}, \eqref{XiExpansion}, and \eqref{normalizedMicroResponse}. The companion \nolinkurl{verification_checks.py} script evaluates the same identities independently at numerical test points.

\section{Consistency Limits and Physical Interpretation}

Several limits provide useful checks:

\begin{enumerate}
\item \textbf{GR limit.} For $f=g=1$, one has $\tau=t$, $b=a$, $\bar\mu=\mu$, and $\mathcal H=H$, so all equations reduce exactly to the classical McVittie relations.

\item \textbf{FLRW limit.} For $m=0$, $\bar\chi=1$ and the metric becomes flat FLRW with lapse $1/f$ and scale factor $b=a/g$. Equations \eqref{rainF1tau} and \eqref{rainF2tau} are then the usual Friedmann equations in the proper time $\tau$ and scale factor $b$.

\item \textbf{Constant rainbow functions.} Constant $f$ and $g$ produce constant rescalings of time and the isotropic scale factor, giving the standard McVittie dynamics in rescaled variables.

\item \textbf{Time-dependent functions.} For varying $f(t)$ and $g(t)$, Eq.~\eqref{rainbowMetric} becomes standard McVittie under the definitions \eqref{btaudef}. The Gravity's Rainbow interpretation is carried by the probe-energy dependence $f(E/E_{\rm P})$, $g(E/E_{\rm P})$ together with the characteristic energy history, which assigns different effective geometries to different-energy probes.

\item \textbf{Horizon branches.} For $0<3\sqrt{3}\,m\mathcal H<1$, the black-hole and cosmological branches lie on opposite sides of $3m$. Equations that divide by $3m-\widetilde R_A$ apply away from the merger. The second factor in Eq.~\eqref{clausiusFactor} gives $\kappa_A=\delta Q=0$ and is a degenerate Clausius projection rather than an additional matter-dynamical branch.

\item \textbf{Observational identifiability.} The exact one-history geometry determines $b(\tau)$ and $m$. The parameter $\epsilon$ belongs to the separate operational closure in Eq.~\eqref{HobsRG}. Under the analytic expansion and standard-redshift assumptions, Eq.~\eqref{epsilonMap} constrains only $\alpha_n+n\beta_n$ after $n$ and $E_0$ are fixed; it cannot infer $f$ and $g$ separately.
\end{enumerate}

\section{Conclusion}

The classical McVittie sector supplies a precise consistency benchmark through its constant central mass, homogeneous density, radially dependent pressure, and vanishing radial momentum density. Applying direct temporal and spatial rainbow factors $f(t)^{-2}$ and $g(t)^{-2}$ while retaining $\mu=m/(2ar)$ produces the areal-radius mass coefficient $m/g(t)$. The factor $(1+\mu)/(1-\mu)$ multiplying $\dot g/g$ in Eq.~\eqref{directExtrinsic} generates the explicit momentum constraint in Eq.~\eqref{naiveMomentum}; hence $m\dot g\neq0$ requires radial energy transport.

A non-accreting perfect-fluid construction follows from $b=a/g$, $d\tau=dt/f$, and $\bar\mu=m/(2br)$. The relation $\dot{\bar\mu}/\bar\mu=\dot g/g-H$ cancels the radial factor and gives the homogeneous extrinsic curvature $K^i{}_j=-\mathcal H\delta^i{}_j$, with $\mathcal H=f(H-\dot g/g)$. In these variables the geometry is exactly McVittie. This result simultaneously establishes algebraic consistency and exposes the identifiability limit: one prescribed energy history contains a reparametrized standard geometry. The Friedmann equations, apparent-horizon condition, surface gravity, entropy, Misner--Sharp energy, and energy-supply vector form a closed system. Projecting the unified first law onto a nondegenerate trapping horizon reconstructs the dynamical Friedmann equation. The alternative factor gives zero surface gravity and zero projected heat flux, while local conservation integrates the dynamical equation to the first Friedmann equation with an additive vacuum-energy constant.

The phenomenological closure connects a common effective redshift response to the measured expansion history under explicit assumptions about redshift, photon propagation, distance duality, and cross-probe universality. For analytic rainbow functions and standard energy redshifting, the leading map $\epsilon=n(\alpha_n+n\beta_n)(E_0/E_{\rm P})^n$ identifies the precise microscopic combination tested by the local slope while retaining the degeneracy between $f$ and $g$. The 32-point cosmic-chronometer compilation and the full-covariance Pantheon+ shape likelihood give $\epsilon=0.0106^{+0.0234}_{-0.0231}$, $H_0=68.75\pm2.63~\mathrm{km\,s^{-1}\,Mpc^{-1}}$, and $\Omega_m=0.3157\pm0.0070$ when the base-$\Lambda$CDM Planck matter-density prior is adopted conditionally. The minimum statistic improves by $0.204$, while AIC and the approximate BIC favor the nested $\epsilon=0$ closure. The fitted amplitude is statistically aligned with zero. Dimensionless figures make the horizon domains, strong-field momentum source, thermodynamic zero, observational envelope, and higher-order response curvature explicit. The rigorous interpretation remains a bound on the stated late-time response, while the microscopic order, probe energy, chromatic propagation, and separate rainbow functions remain model inputs. The McVittie mass dependence resides in the pressure profile, trapping horizons, surface gravity, and local geometry, opening a complementary program based on lensing, time delays, turnaround scales, and energy-resolved strong-field propagation.


\begin{thebibliography}{99}

\bibitem{friedmann1922} A. Friedmann, ``Uber die Krummung des Raumes,'' Z. Phys. \textbf{10}, 377--386 (1922).

\bibitem{friedmann1924} A. Friedmann, ``On the possibility of a world with constant negative curvature,'' Z. Phys. \textbf{21}, 326--332 (1924).

\bibitem{lemaitre1927} G. Lemaitre, ``A homogeneous universe of constant mass and increasing radius,'' Ann. Soc. Sci. Bruxelles A \textbf{47}, 49--59 (1927).

\bibitem{mcVittie1933} G. C. McVittie, ``The mass-particle in an expanding universe,'' Mon. Not. R. Astron. Soc. \textbf{93}, 325--339 (1933).

\bibitem{nolan1998} B. C. Nolan, ``A point mass in an isotropic universe: Existence, uniqueness, and basic properties,'' Phys. Rev. D \textbf{58}, 064006 (1998), arXiv:gr-qc/9805041.

\bibitem{kaloper2010} N. Kaloper, M. Kleban, and D. Martin, ``McVittie's legacy: Black holes in an expanding universe,'' Phys. Rev. D \textbf{81}, 104044 (2010), arXiv:1003.4777.

\bibitem{faraoni2012} V. Faraoni, A. F. Zambrano Moreno, and R. Nandra, ``Making sense of the bizarre behaviour of horizons in the McVittie spacetime,'' Phys. Rev. D \textbf{85}, 083526 (2012), arXiv:1202.0719.

\bibitem{jacobson1995} T. Jacobson, ``Thermodynamics of spacetime: The Einstein equation of state,'' Phys. Rev. Lett. \textbf{75}, 1260--1263 (1995), arXiv:gr-qc/9504004.

\bibitem{magueijo2004} J. Magueijo and L. Smolin, ``Gravity's Rainbow,'' Class. Quantum Grav. \textbf{21}, 1725--1736 (2004), arXiv:gr-qc/0305055, doi:10.1088/0264-9381/21/7/001.

\bibitem{galan2006} P. Galan and G. A. Mena Marugan, ``Entropy and temperature of black holes in a gravity's rainbow,'' Phys. Rev. D \textbf{74}, 044035 (2006), arXiv:gr-qc/0608061, doi:10.1103/PhysRevD.74.044035.

\bibitem{ling2007} Y. Ling, ``Rainbow universe,'' JCAP \textbf{08} (2007) 017, arXiv:gr-qc/0609129, doi:10.1088/1475-7516/2007/08/017.

\bibitem{lingli2007} Y. Ling, X. Li, and H. Zhang, ``Thermodynamics of modified black holes from gravity's rainbow,'' Mod. Phys. Lett. A \textbf{22}, 2749--2756 (2007), arXiv:gr-qc/0512084, doi:10.1142/S0217732307022931.

\bibitem{awad2013} A. Awad, A. F. Ali, and B. Majumder, ``Nonsingular rainbow universes,'' JCAP \textbf{10} (2013) 052, arXiv:1308.4343, doi:10.1088/1475-7516/2013/10/052.

\bibitem{amelino2013} G. Amelino-Camelia, M. Arzano, G. Gubitosi, and J. Magueijo, ``Rainbow gravity and scale-invariant fluctuations,'' Phys. Rev. D \textbf{88}, 041303(R) (2013), arXiv:1307.0745, doi:10.1103/PhysRevD.88.041303.

\bibitem{ali2014} A. F. Ali, ``Black hole remnant from gravity's rainbow,'' Phys. Rev. D \textbf{89}, 104040 (2014), arXiv:1402.5320, doi:10.1103/PhysRevD.89.104040.

\bibitem{aliblackrings2014} A. F. Ali, M. Faizal, and M. M. Khalil, ``Remnants of black rings from gravity's rainbow,'' JHEP \textbf{12} (2014) 159, arXiv:1409.5745, doi:10.1007/JHEP12(2014)159.

\bibitem{alifaizal2015remnant} A. F. Ali, M. Faizal, and M. M. Khalil, ``Remnant for all black objects due to gravity's rainbow,'' Nucl. Phys. B \textbf{894}, 341--360 (2015), arXiv:1410.5706, doi:10.1016/j.nuclphysb.2015.03.014.

\bibitem{alifaizal2015lhc} A. F. Ali, M. Faizal, and M. M. Khalil, ``Absence of black holes at LHC due to gravity's rainbow,'' Phys. Lett. B \textbf{743}, 295--300 (2015), arXiv:1410.4765, doi:10.1016/j.physletb.2015.02.065.

\bibitem{garattini2015} R. Garattini and E. N. Saridakis, ``Gravity's Rainbow: a bridge towards Ho\v{r}ava--Lifshitz gravity,'' Eur. Phys. J. C \textbf{75}, 343 (2015), arXiv:1411.7257, doi:10.1140/epjc/s10052-015-3562-y.

\bibitem{ashour2016} A. Ashour, M. Faizal, A. F. Ali, and F. Hammad, ``Branes in Gravity's Rainbow,'' Eur. Phys. J. C \textbf{76}, 264 (2016), arXiv:1602.04926, doi:10.1140/epjc/s10052-016-4124-7.

\bibitem{hendi2016gb} S. H. Hendi, M. Momennia, B. Eslam Panah, and M. Faizal, ``Nonsingular universes in Gauss--Bonnet Gravity's Rainbow,'' Astrophys. J. \textbf{827}, 153 (2016), arXiv:1703.00480, doi:10.3847/0004-637X/827/2/153.

\bibitem{hendi2016tov} S. H. Hendi, G. H. Bordbar, B. Eslam Panah, and S. Panahiyan, ``Modified TOV in gravity's rainbow: properties of neutron stars and dynamical stability conditions,'' JCAP \textbf{09} (2016) 013, arXiv:1509.05145, doi:10.1088/1475-7516/2016/09/013.

\bibitem{gorji2017} M. A. Gorji, K. Nozari, and B. Vakili, ``Gravity's rainbow: A bridge between LQC and DSR,'' Phys. Lett. B \textbf{765}, 113--119 (2017), arXiv:1606.00910, doi:10.1016/j.physletb.2016.12.023.

\bibitem{feng2017} Z.-W. Feng and S.-Z. Yang, ``Thermodynamic phase transition of a black hole in rainbow gravity,'' Phys. Lett. B \textbf{772}, 737--742 (2017), arXiv:1708.06627, doi:10.1016/j.physletb.2017.07.057.

\bibitem{hayward1996} S. A. Hayward, ``Gravitational energy in spherical symmetry,'' Phys. Rev. D \textbf{53}, 1938--1949 (1996), arXiv:gr-qc/9408002.

\bibitem{hayward1998} S. A. Hayward, ``Unified first law of black-hole dynamics and relativistic thermodynamics,'' Class. Quantum Grav. \textbf{15}, 3147--3162 (1998), arXiv:gr-qc/9710089.

\bibitem{jiang2011} K.-X. Jiang, S.-M. Ke, and D.-T. Peng, ``Hawking radiation as tunneling and the unified first law of thermodynamics for a class of dynamical black holes,'' Int. J. Mod. Phys. D \textbf{18}, 1707--1717 (2009), arXiv:1105.0595.

\bibitem{akbar2017} M. Akbar, T. Brahimi, and S. M. Qaisar, ``Thermodynamic analysis of cosmological black hole,'' Commun. Theor. Phys. \textbf{67}, 47 (2017).

\bibitem{abdusattar2022} H. Abdusattar, S.-B. Kong, Y. Yin, and Y.-P. Hu, ``The Hawking--Page-like phase transition from FRW spacetime to McVittie black hole,'' JCAP \textbf{08} (2022) 060, arXiv:2203.10868.


\bibitem{moresco2020} M. Moresco, R. Jimenez, L. Verde, A. Cimatti, and L. Pozzetti, ``Setting the Stage for Cosmic Chronometers. II. Impact of Stellar Population Synthesis Models Systematics and Full Covariance Matrix,'' Astrophys. J. \textbf{898}, 82 (2020), arXiv:2003.07362, doi:10.3847/1538-4357/ab9eb0.

\bibitem{favale2023} A. Favale, A. G\'omez-Valent, and M. Migliaccio, ``Cosmic chronometers to calibrate the ladders and measure the curvature of the Universe. A model-independent study,'' Mon. Not. R. Astron. Soc. \textbf{523}, 3406--3422 (2023), arXiv:2301.09591, doi:10.1093/mnras/stad1621.

\bibitem{scolnic2022} D. Scolnic et al., ``The Pantheon+ Analysis: The Full Data Set and Light-curve Release,'' Astrophys. J. \textbf{938}, 113 (2022), arXiv:2112.03863, doi:10.3847/1538-4357/ac8b7a.

\bibitem{brout2022} D. Brout et al., ``The Pantheon+ Analysis: Cosmological Constraints,'' Astrophys. J. \textbf{938}, 110 (2022), arXiv:2202.04077, doi:10.3847/1538-4357/ac8e04.

\bibitem{planck2018} N. Aghanim et al. [Planck Collaboration], ``Planck 2018 results. VI. Cosmological parameters,'' Astron. Astrophys. \textbf{641}, A6 (2020), arXiv:1807.06209.

\bibitem{etherington1933} I. M. H. Etherington, ``On the definition of distance in general relativity,'' Philos. Mag. \textbf{15}, 761--773 (1933); reprinted in Gen. Relativ. Gravit. \textbf{39}, 1055--1067 (2007), doi:10.1007/s10714-007-0447-x.

\end{thebibliography}
\end{document}